\documentclass[prl,aps,reprint]{revtex4-1}

\usepackage{amsmath,amssymb,graphicx,color}
\usepackage{soul} 

\soulregister\ref{7}
\soulregister\eqref{7}
\soulregister\cite{7}
\soulregister\onlinecite{7}

\begin{document}

\title{Circularly polarized states spawning from bound states in the continuum}

\author{Wenzhe Liu$^{1,2}$}
\author{Bo Wang$^{1,2}$}
\author{Yiwen Zhang$^{1,2}$}
\author{Jiajun Wang$^{1,2}$}
\author{Maoxiong Zhao$^{1,2}$}
\author{Fang Guan$^{1,2}$}
\author{Xiaohan Liu$^{1,2}$}
\author{Lei Shi$^{1,2}$}
\email{lshi@fudan.edu.cn}
\author{Jian Zi$^{1,2}$}
\email{jzi@fudan.edu.cn}
\affiliation{$^{1}$ State Key Laboratory of Surface Physics, Key Laboratory of Micro- and Nano-Photonic Structures (Ministry of Education) and Department of Physics, Fudan University, Shanghai 200433, China}
\affiliation{$^{2}$ Collaborative Innovation Center of Advanced Microstructures, Fudan University, Shanghai 200433, China}

\begin{abstract}
Bound states in the continuum in periodic photonic systems like photonic crystal slabs are proved to be accompanied by vortex polarization singularities on the photonic bands in the momentum space. The winding structures of polarization states not only widen the field of topological physics but also show great potential that such systems could be applied in polarization manipulating. In this work, we report the phenomenon that by in-plane inversion ($C_2$) symmetry breaking, pairs of circularly polarized states could spawn from the eliminated Bound states in the continuum. Along with the appearance of the circularly polarized states as the two poles of the Poincar\'e sphere together with linearly polarized states covering the equator, full coverage on the Poincar\'e sphere could be realized. As an application, ellipticity modulation of linear polarization is demonstrated in the visible frequency range. This phenomenon provides new degree of freedom in modulating polarization.
\end{abstract}

\maketitle


Polarization is one of electromagnetic wave's most essential properties. Controlling the polarization is found very important in a lot of fields, such as 3d imaging~\cite{pircher2004transversal}, optical communication~\cite{keiser2003optical}, and quantum optics~\cite{mandel1995optical}. In recent years, great attention has been paid to modulate polarization of light with compact structures such as metasurfaces~\cite{zhao2011manipulating, glybovski2016metasurfaces, kruk2016invited, kruk2017functional} instead of classical wave plates and polarizers, which are more applicable in on-chip devices. On the other hand, application of another kind of structure, photonic crystal (PhC) slab, in modulating polarization, is capturing interest nowadays~\cite{lobanov2015polarization, hsu2017polarization, guo2017topologically, guo2019arbitrary}. Their simplicity in fabrication, designable band structures, and complex polarization features in the momentum space which are topologically linked with bound states in the continuum~\cite{hsu2013observation, zhen2014topological, yang2014analytical, guo2017topologically, bulgakov2017bound, zhang2018observation, doeleman2018experimental, song2018cherenkov, dai2018topologically, he2018toroidal, jin2018topologically, koshelev2018asymmetric, guo2019arbitrary, chen2019singularities, cerjan2019bound, koshelev2019nonradiating, sadrieva2019multipolar} are favorable for polarization modulation. However, the reported PhC slabs only support nearly linearly polarized resonances which only cover a small area on the Poincar\'e sphere. The coverage on the Poincar\'e characterized the polarization properties of the system. Small coverage with the two poles (circular states of polarization) missing indicates its limited capability in full Stokes polarization modulating.
In this paper, we report that, by breaking the in-plane inversion ($C_2$) symmetry of a PhC slab, the at-$\mathrm{\Gamma}$ BICs also known as vortex polarization singularities (V-points)~\cite{hsu2013observation, zhen2014topological, bulgakov2017bound, zhou2018observation, zhang2018observation, doeleman2018experimental, jin2018topologically, chen2019singularities} on photonic bands will be eliminated~\cite{koshelev2018asymmetric}. As the singularities are broken, the winding of the main axis of the polarization states are preserved, leading to generation of pairs of circularly polarized states (C-points) near the $\mathrm{\Gamma}$ point. With a line of linearly polarized states (L-line) enclosing the position of the original BIC, the generation of C-points would even enable full coverage on the Poincar\'e sphere. We then demonstrated the application of this phenomenon in ellipticity modulation of light.

\begin{figure}[t]
\centering
\includegraphics[scale=1]{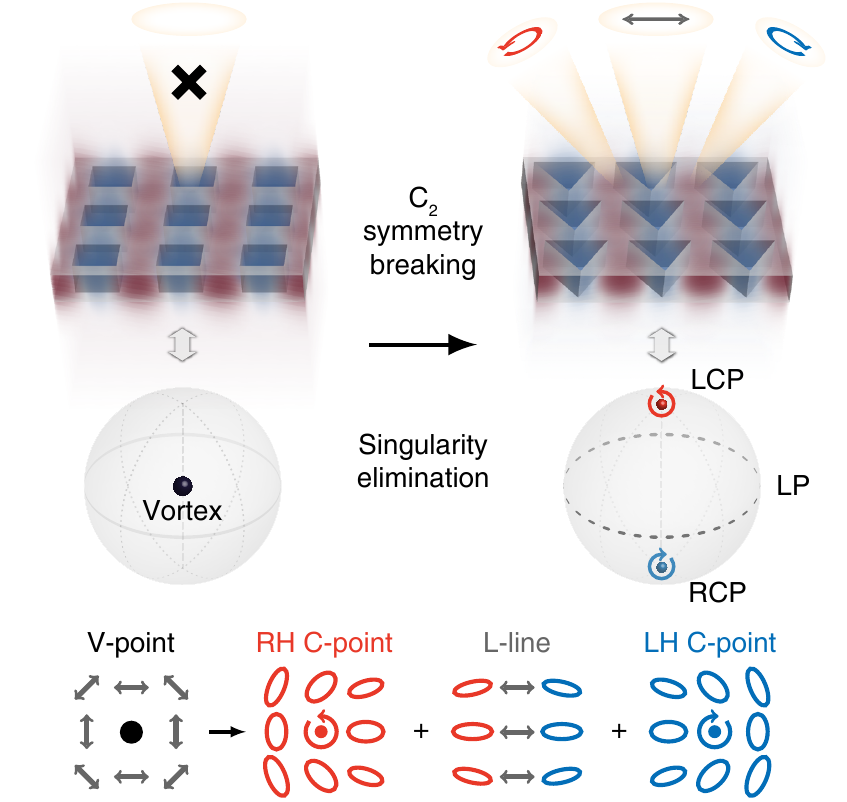}
\caption{Schematic view of how circularly polarized states (C-points) spawn from a bound state in the continuum when in-plane inversion ($C_2$) symmetry is broken. C-points and lines of linearly polarized states (L-lines) could be found near $\mathrm{\Gamma}$ point. The asymmetric field of the mode in the $C_2$ symmetry broken structure induces extra in-plane multipole moments. L(R)CP: left(right)-handed circular polarization; L(R)H: left(right)-handed.}
\label{fig:1}
\end{figure}
\begin{figure*}[t]
\centering
\includegraphics[scale=1]{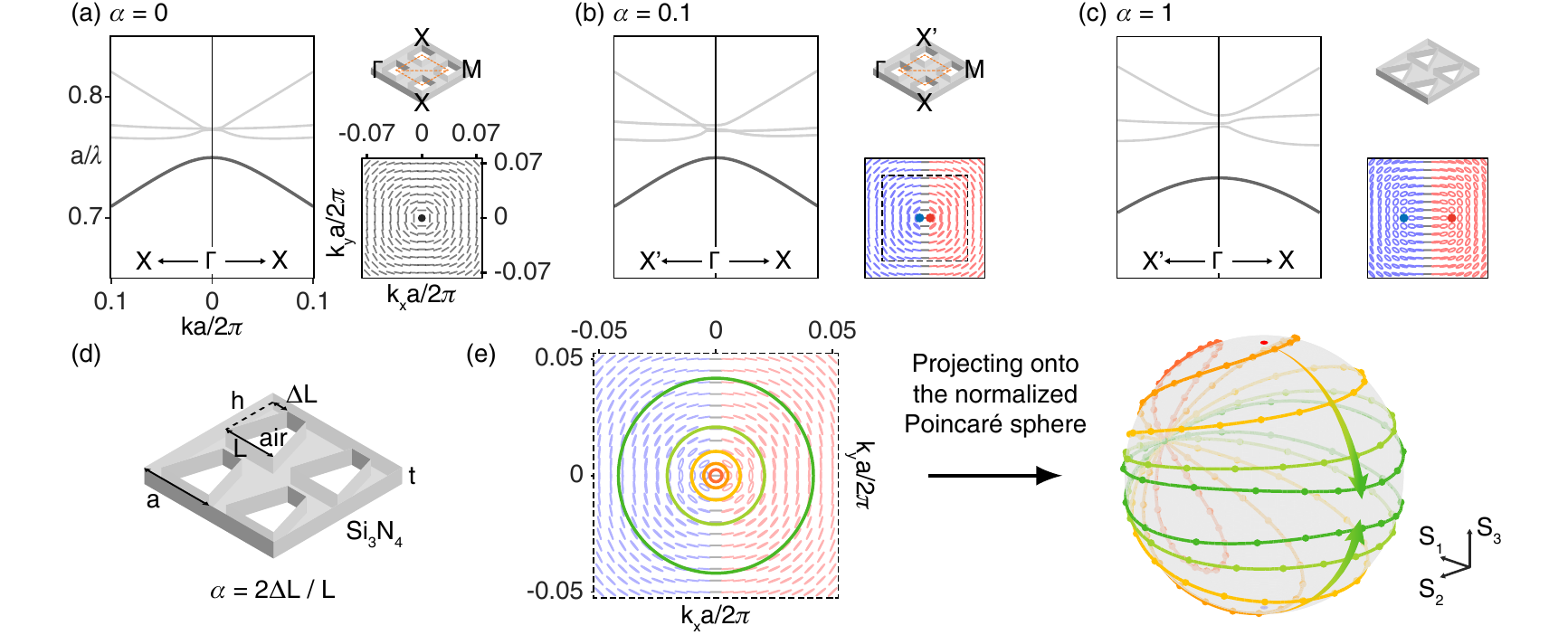}
\caption{(a-c) The TE-like band structures and polarization maps in the vicinity of the $\mathrm{\Gamma}$ point of structures with different values of $\mathrm{\alpha}$. We focus on band TE2, which is marked with bold dark line on the band structure diagram. On the polarization maps, the polarization states are represented by ellipses of which the red color corresponds to left-handed states and the blue color corresponds to right-handed ones. The zoomed-in map of the enclosed area in (b) will be shown in (e). (d) A schematic view of the structure and the definition of the asymmetry parameter $\alpha$. (e) The circular loops of polarization states mapped from the Brillouin zone (BZ) to the Poincar\'e sphere. The radius of the loop are 0.0025, 0.005, 0.01, 0.02 and 0.04 from orange to green. The data points are sampled per 10 degrees on every loop in the BZ. }
\label{fig:2}
\end{figure*}

For a 2D PhC slab, there would be a series of Bloch resonant modes with different frequencies $f$ and wave vectors $k$ forming photonic bands. Radiation polarization states of these modes on an arbitrary band could be projected into the structure plane and mapped onto the Brillouin zone (BZ), which defines a polarization field in the momentum space~\cite{hsu2013observation, zhen2014topological, bulgakov2017bound, zhou2018observation, zhang2018observation, doeleman2018experimental, jin2018topologically, chen2019singularities}. These Bloch modes are mostly radiative, unless destructive interference or symmetry mismatch makes them non-radiative, i.e. BICs. The BICs appear as vortex singularities (V-points) in the polarization field. To characterize the polarization properties of the system, we can map the polarization states of the Bloch modes onto the Poincar\'e sphere of which the coordinates are specified by Stokes parameters $S_0$, $S_1$, $S_2$ and $S_3$. In this language, those non-radiative singularities will have zero values of Stokes parameters. Thus, they are also singularities in the normalized Stokes parametric space, shown as the left panel of Fig. \ref{fig:1}. Proved by C. W. Hsu et al.~\cite{hsu2017polarization}, the polarization states of radiative modes are almost linear in the whole BZ of a 2D PhC Slab with $C_2$ symmetry, of which the low ellipticity is a result from the perturbative non-Hermiticity in the system. As a consequence, the corresponding polarization map projected from the BZ onto the normalized Poincar\'e sphere should be a belt near the equator plus a singularity. Since a large area including the two poles is not covered on the Poincar\'e sphere, it will limit the applications of 2D PhC Slabs such as polarization modulating. Recently, it was known that the symmetry-protected BICs at the $\mathrm{\Gamma}$ point origin from multipole moment perpendicular to the structure plane in a $C_2$ symmetric system~\cite{doeleman2018experimental, koshelev2018asymmetric, chen2019singularities, sadrieva2019multipolar}.
If $C_2$ symmetry of the system is slightly broken, the in-plane electromagnetic field will be perturbed as shown in Fig. \ref{fig:1}, and non-zero multipole moments lying in the structure plane would be induced.
As researched in singular optics~\cite{nye1999natural, berry2001polarization, dennis2002polarization, freund2002polarization, schoonover2006polarization, burresi2009observation, song2018valley, bliokh2019geometric}, V-points ($S_0,S_1,S_2,S_3=0$) result from the collision of C-points ($S_1,S_2=0$) and L-lines ($S_3=0$)~\cite{schoonover2006polarization, d2017topological, otte2018polarization}. Thus with the perturbation induced by $C_2$ symmetry breaking, the integer-charged V-points at the $\mathrm{\Gamma}$ point will decompose into pairs of half-charged C-points and L-lines, as illustrated in the right panel of Fig. \ref{fig:1}. This would break the pre-mentioned limitation on ellipticity.

%
To verify the physical picture above, we designed freestanding 2D PhC slabs and studied the phenomenon in theory and experiment. The slabs here are made of silicon nitride ($\mathrm{Si_3N_4}$), of which the refractive index is about 2.02. The thickness $t$ of the slabs are chosen to be 100 nm. Square lattices of holes with a period $a$ of 450 nm are etched on the slabs with the original shape square. Then, to break the $C_2$ symmetry, the squares are transformed to isosceles trapezoids with their area $S = S_0 = 247\mathrm{\ nm} \times 247\mathrm{\ nm}$ unchanged. During the transformation, the height $h$ and the baseline length $L$ of any trapezoid are maintained equal ($L = h$). The asymmetry parameter $\alpha$ here is then defined as the ratio of the reduced topline length $2 \Delta L$ to the baseline length $L$, shown in Fig. \ref{fig:2}(d). Increasing $\alpha$ from 0 to 1 would transform the holes from squares to isosceles triangles, while $L = \sqrt{2 S_0 / (2 -\alpha)}$.

\begin{figure}[tb]
\centering
\includegraphics[scale=1]{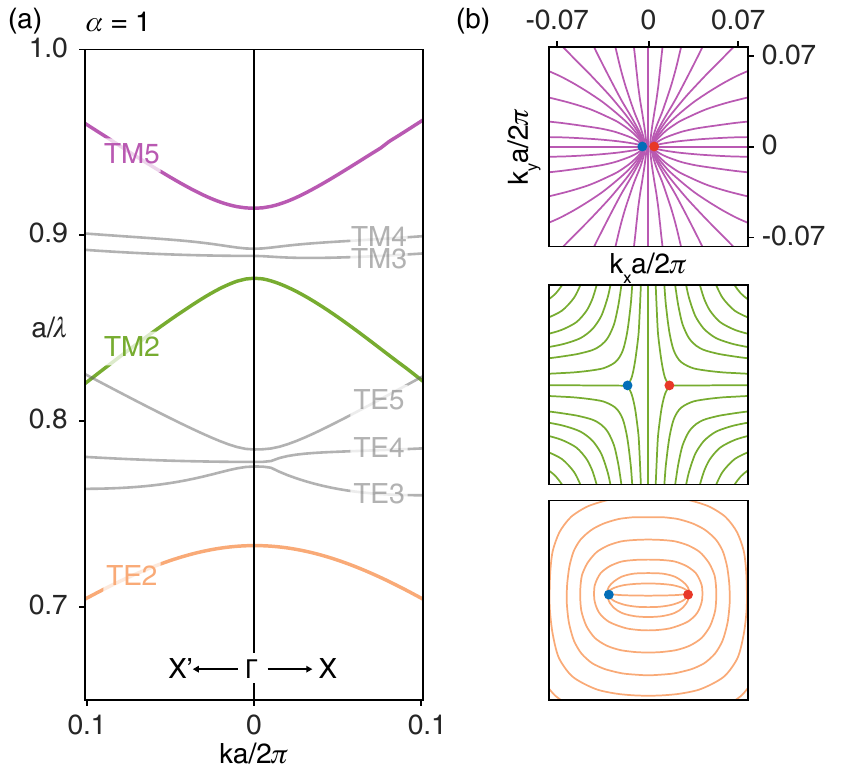}
\caption{(a) The band structures along $\mathrm{\Gamma}$-X \& $\mathrm{\Gamma}$-X' near the $\mathrm{\Gamma}$ point. (b) Different configurations of polarization main axis around the C-points on the bands TE2(bottom, `lemon'), TM2(middle, `star'), TM5(top, `monstar') of the structure with $\mathrm{\alpha} = 1$. }
\label{fig:3}
\end{figure}

\begin{figure}[b]
\centering
\includegraphics[scale=1]{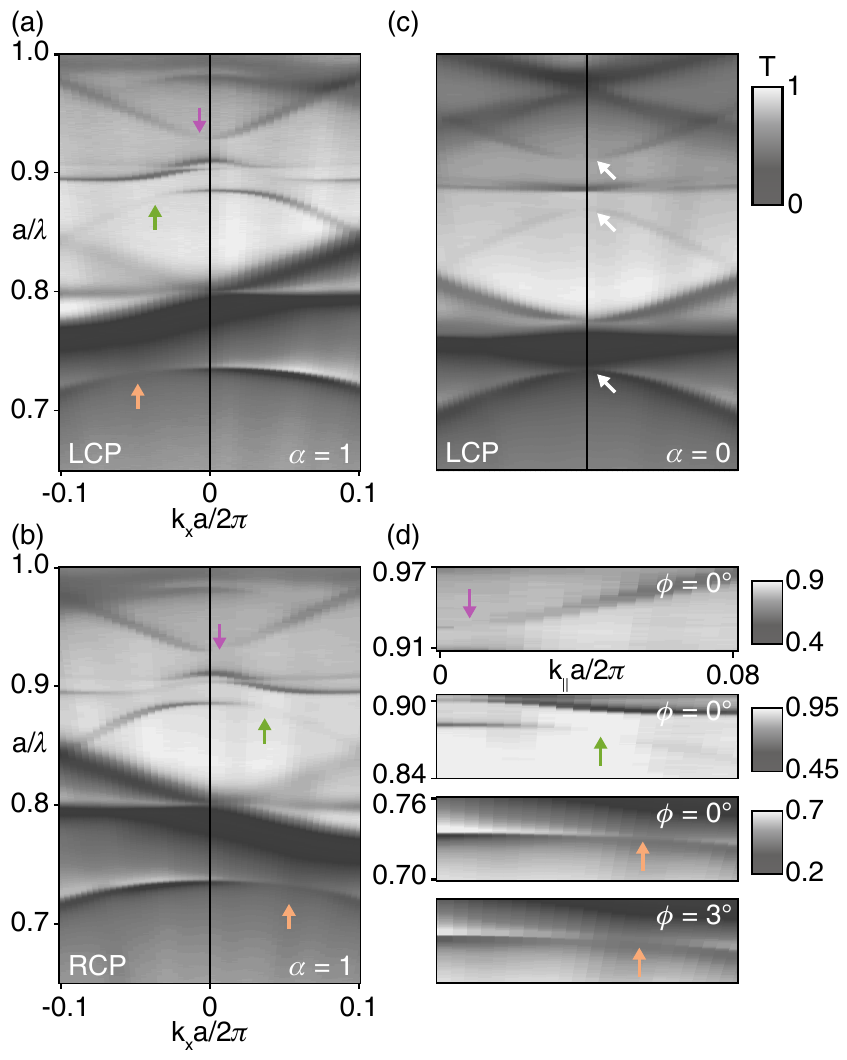}
\caption{The measured angle-resolved transmittance spectra in the visible frequency range under circularly polarized incidence. (a) The spectra of the $\mathrm{\alpha} = 1$ sample, LCP incidence. Vanished regions are pointed out by arrows, implying the existence of C-points after breaking the $C_2$ symmetry. The colors correspond to the colored bands in Fig.\ref{fig:3}. (b) The spectra of the $\mathrm{\alpha} = 1$ sample, RCP incidence. (c) The spectra of the $\mathrm{\alpha} = 0$ sample, LCP incidence. The vanished regions marked by the white arrows are the BICs. (d) The zoomed-in plots of the vanished regions in (c). Note that the C-points on band TE2 are rather in the direction $\phi\sim\pm3^\circ$ to $\mathrm{\Gamma}$-X than in the direction along $\mathrm{\Gamma}$-X, so we plotted the spectra along this direction.}
\label{fig:4}
\end{figure}

We first focus on the BIC at the $\mathrm{\Gamma}$ point of the band TE2 for example. Applying finite element method calculation, we firstly confirmed that the quality factors of the states at the $\mathrm{\Gamma}$ point on band TE2 obey the law $\alpha^{-2}$ when the perturbation is small, which is proved in~\cite{koshelev2018asymmetric}. The results could be found in the Supplemental Material (SM)~\cite{sm}. Subsequently, we obtained the projected radiation polarization states $(d_x, d_y)$ on band TE2 to see the generation of C-points. The band structures and polarization maps near the $\mathrm{\Gamma}$ point of systems with different asymmetry parameters $\alpha = 0, 0.1 \& 1$ are plotted in Fig. \ref{fig:2}(a-c). We can clearly see in Fig. \ref{fig:2}(a) that when $\alpha = 0$, the polarization states are close to linear, and there would be a V-point (BIC) at the $\mathrm{\Gamma}$ point. When $\alpha$ slightly increases to 0.1, shown in Fig. \ref{fig:2}(b), the at-$\mathrm{\Gamma}$ vortex is broken and a pair of C-points with opposite chirality (the two poles) appear very close to the $\mathrm{\Gamma}$ point, clamping an L-line. This indicates that the areas around the two poles of the Poincar\'e sphere are covered. To be noted that, we have an L-line far away from the $\mathrm{\Gamma}$ point enclosing the original at-$\mathrm{\Gamma}$ V-point (BIC) in our system, shown in the SM~\cite{sm}. This L-line corresponds to the entire equator of the Poincar\'e sphere, and due to continuity, the full Poincar\'e sphere is thus covered. To prove this, we projected circular loops in the BZ in the vicinity of the $\mathrm{\Gamma}$ point onto the normalized Poincar\'e sphere under condition that $\alpha$ equals to 0.1. We find that the larger area the loop in the BZ encloses [shown in Fig. \ref{fig:2}(e), left panel], the closer the loop on the Poincar\'e sphere is to the equator [shown in Fig. \ref{fig:2}(e), right panel]. We can see that the coverage approaching the full Poincar\'e sphere. Note that there is a point on the equator pinned on every loop, which corresponds to the clamped L-line protected by mirror symmetry in our system. As the perturbation is larger [$\alpha = 1$, shown in Fig. \ref{fig:2}(c)], the C-points will move away from the original position of the vortex. A larger distance between C-points would make the experimental observation more easy for us.

Furthermore, this phenomenon induced by $C_2$ symmetry breaking is promised to be found on any band. We analyzed the bands TE2, TM2 and TM5 of the above-analyzed 2D PhC slab with $\mathrm{\alpha} = 1$, finding the phenomenon not only common but also giving different topological polarization configurations of C-points near the $\mathrm{\Gamma}$ point~\cite{nye1999natural, berry2001polarization, dennis2002polarization, freund2002polarization, schoonover2006polarization, burresi2009observation}. A `lemon' is found on TE2, a `star' is found on TM2, and a `monstar' is found on TM5. The band structures along $\mathrm{\Gamma}$-X and $\mathrm{\Gamma}$-X' are plotted in Fig. \ref{fig:3}(a), while the polarization distributions of the different C-points are plotted in Fig. \ref{fig:3}(b). The different configurations on different bands give us much more freedom in designing the eigenmodes with desired polarization states at desired incident or radiating directions.

\begin{figure}[t]
\centering
\includegraphics[scale=1]{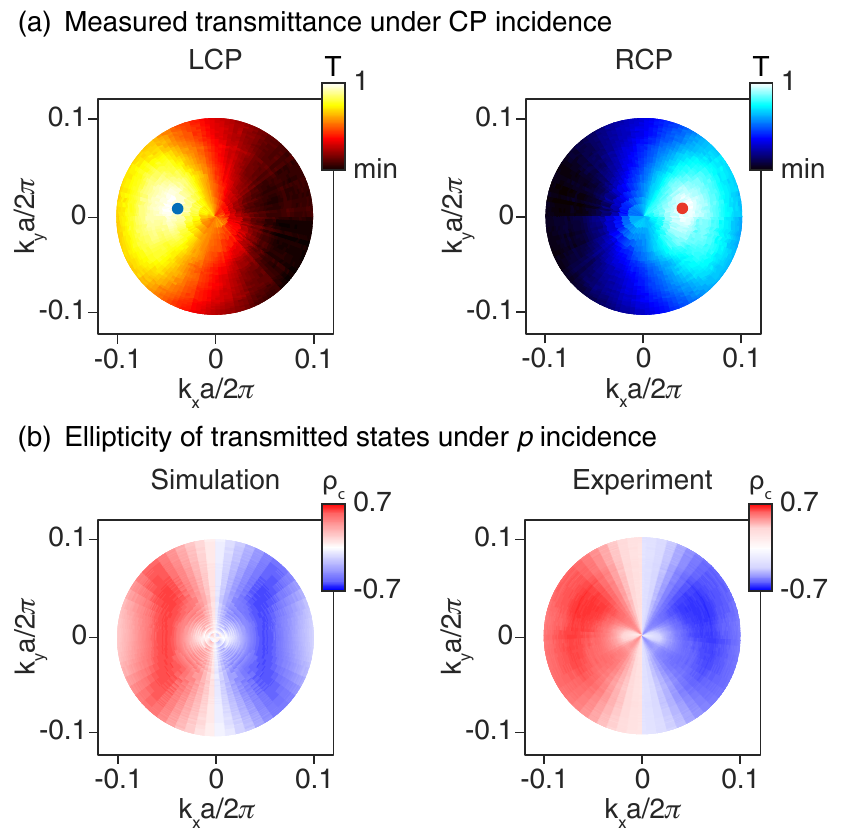}
\caption{(a) Experimental observation of the coverage on Poincar\'e sphere presented in the form of transmittance maps along the band TE2 in the visible frequency range. (b) Simulated and measured ellipticity maps of transmitted polarization states under $p$ incidence on the same band as in (a).}
\label{fig:5}
\end{figure}

For an experimental observation of this phenomenon in the visible frequency range, we fabricated the simulated PhC slabs. The designed hole array is etched on a freestanding silicon nitride layer applying reactive-ion etching with the assistance of the mask poly(methyl methacrylate) layer etched by the electron beam lithography. For optical measurements, we applied our home-made momentum-space imaging spectroscopy system~\cite{zhang2018observation} to obtain their polarization-dependent angle-resolved transmittance spectra. Under LCP (and RCP) incidence, the spectra [Fig. \ref{fig:4}(c)] of the $\alpha = 0$ sample shows no diminishing point other than the non-excited BICs at the $\mathrm{\Gamma}$ point. In contrast, the at-$\mathrm{\Gamma}$ states of the $\mathrm{\alpha} = 1$ sample [Fig. \ref{fig:4}(a)] are radiative and some regions near the $\mathrm{\Gamma}$ point on the bands are diminished instead under LCP incidence. The non-excited states in these regions are hence RH polarized with C-points at the center. Changing the incidence to RCP, the diminished regions of the $\alpha = 1$ sample switch to the other side of the BZ, meaning another C-point with opposite chirality [Fig. \ref{fig:4}(b)]. The zoomed-in plots of the diminished regions are shown in Fig. \ref{fig:4}(d), with the detailed RCP spectra showing the C-point on band TE2 included in the SM~\cite{sm}. The existence of the C-points are proved and agree with the simulations above, proving the phenomenon.

Basing on the angle-resolved polarization dependent spectra we measured, we mapped the transmittance along the band TE2 as a function of $f$ and $k$ under LCP and RCP incidence in order to observe the C-points and the coverage on Poincar\'e sphere more clearly, plotted in Fig. \ref{fig:5}(a). Note that, the maps not only cover the BZ, but also span a spectral range. Changing the period to 400 nm, the bottom line length of the triangle to 300 nm, the thickness of the slab to 215 nm, we were able to suppress the effect of Fano resonances, and the reflectance (1 - transmittance) in such a case shall approximately correspond to the inner product of the projected polarization state and the incident circular polarization [detailed in the SM~\cite{sm}]. On either map, we can see one spot where the transmittance is close to 1, indicating a C-point with which the incident light cannot couple. From the map, the distribution of polarization states is clearly exhibited, and we can see that the coverage approaching full Poincar\'e sphere is directly observed in a rather small range of the BZ.

The coverage approaching full Poincar\'e sphere gives us wider choices in modulating polarization with PhC slabs. Here in the visible frequency range, we demonstrate the application in modulating the ellipticity $\rho_c = S_3/S_0$. ($S_0,S_3$ are the first and the fourth Stokes parameters.) In Fig. \ref{fig:5}(b), we plotted the ellipticity map of the completely polarized transmitted light on resonances of the band TE2, applying the sample measured in Fig. \ref{fig:5}(a). $\rho_c$ here is calculated from the on-resonance transmittance measured with $p$-, $s$- and $\pm45^\circ$-polarized analyzers. As we can see from Fig. \ref{fig:5}(b), the linearly $p$-polarized incident light ($\rho_c = 0$) is transformed to elliptically polarized light, of which $\rho_c$ reaches 0.7 at most. The results of experiments (right panel) match up with rigorous coupled wave algorithm (RCWA) simulations (left panel) well. Choosing the right parameters and with higher fabrication precision, the linearly polarized light can be perfectly transformed into circularly polarized light in theory. An example is detailed in the SM~\cite{sm}. To be noted that, this effect can be viewed as a result of strong orbit-spin coupling induced by $C_2$ symmetry breaking, and similar phenomenon in a mono-chromatic beam also induced by symmetry breaking has been shown previously~\cite{yin2013photonic, bliokh2015spin}.

In conclusion, we reported the phenomenon that by breaking the in-plane inversion symmetry ($C_2$) of a 2D PhC slab, the non-radiative vortex singularities at the $\mathrm{\Gamma}$ point will be eliminated. As a consequence, C-points with different polarization configuration could be found near the $\mathrm{\Gamma}$ point. Together with an L-line looping around, the full Poincar\'e sphere coverage can be realized. We theoretically verified and experimentally observed the generation of C-points and the coverage approaching full Poincar\'e sphere. This phenomenon enriches the knowledge on polarization properties of 2D PhC slabs in the momentum space. The coverage on the Poincar\'e sphere offers us new latitude to modulate polarization with 2D PhC slabs, which may be applied in vector beam generating or quantum optics.

We thank Dr. Ang Chen, Prof. Chao Peng, Prof. Chia Wei Hsu, Prof. Dezhuan Han, Prof. Ling Lu, Prof. Meng Xiao, Dr. Haiwei Yin and Prof. Shaoyu Yin for helpful discussions. The work was supported by 973 Program and China National Key Basic Research Program (2015CB659400, 2016YFA0301100, 2016YFA0302000 and 2018YFA0306201) and National Science Foundation of China (11774063, 11727811, 91750102 and 11604355). The research of L. S. was further supported by Science and Technology Commission of Shanghai Municipality (17ZR1442300, 17142200100).

W. L., B. W., and Y. Z. contributed equally to this work.


\begin{thebibliography}{40}%
\makeatletter
\providecommand \@ifxundefined [1]{%
 \@ifx{#1\undefined}
}%
\providecommand \@ifnum [1]{%
 \ifnum #1\expandafter \@firstoftwo
 \else \expandafter \@secondoftwo
 \fi
}%
\providecommand \@ifx [1]{%
 \ifx #1\expandafter \@firstoftwo
 \else \expandafter \@secondoftwo
 \fi
}%
\providecommand \natexlab [1]{#1}%
\providecommand \enquote  [1]{``#1''}%
\providecommand \bibnamefont  [1]{#1}%
\providecommand \bibfnamefont [1]{#1}%
\providecommand \citenamefont [1]{#1}%
\providecommand \href@noop [0]{\@secondoftwo}%
\providecommand \href [0]{\begingroup \@sanitize@url \@href}%
\providecommand \@href[1]{\@@startlink{#1}\@@href}%
\providecommand \@@href[1]{\endgroup#1\@@endlink}%
\providecommand \@sanitize@url [0]{\catcode `\\12\catcode `\$12\catcode
  `\&12\catcode `\#12\catcode `\^12\catcode `\_12\catcode `\%12\relax}%
\providecommand \@@startlink[1]{}%
\providecommand \@@endlink[0]{}%
\providecommand \url  [0]{\begingroup\@sanitize@url \@url }%
\providecommand \@url [1]{\endgroup\@href {#1}{\urlprefix }}%
\providecommand \urlprefix  [0]{URL }%
\providecommand \Eprint [0]{\href }%
\providecommand \doibase [0]{http://dx.doi.org/}%
\providecommand \selectlanguage [0]{\@gobble}%
\providecommand \bibinfo  [0]{\@secondoftwo}%
\providecommand \bibfield  [0]{\@secondoftwo}%
\providecommand \translation [1]{[#1]}%
\providecommand \BibitemOpen [0]{}%
\providecommand \bibitemStop [0]{}%
\providecommand \bibitemNoStop [0]{.\EOS\space}%
\providecommand \EOS [0]{\spacefactor3000\relax}%
\providecommand \BibitemShut  [1]{\csname bibitem#1\endcsname}%
\let\auto@bib@innerbib\@empty
\bibitem [{\citenamefont {Pircher}\ \emph {et~al.}(2004)\citenamefont
  {Pircher}, \citenamefont {Goetzinger}, \citenamefont {Leitgeb},\ and\
  \citenamefont {Hitzenberger}}]{pircher2004transversal}%
  \BibitemOpen
  \bibfield  {author} {\bibinfo {author} {\bibfnamefont {M.}~\bibnamefont
  {Pircher}}, \bibinfo {author} {\bibfnamefont {E.}~\bibnamefont {Goetzinger}},
  \bibinfo {author} {\bibfnamefont {R.}~\bibnamefont {Leitgeb}}, \ and\
  \bibinfo {author} {\bibfnamefont {C.~K.}\ \bibnamefont {Hitzenberger}},\
  }\href@noop {} {\bibfield  {journal} {\bibinfo  {journal} {Physics in
  Medicine \& Biology}\ }\textbf {\bibinfo {volume} {49}},\ \bibinfo {pages}
  {1257} (\bibinfo {year} {2004})}\BibitemShut {NoStop}%
\bibitem [{\citenamefont {Keiser}(2003)}]{keiser2003optical}%
  \BibitemOpen
  \bibfield  {author} {\bibinfo {author} {\bibfnamefont {G.}~\bibnamefont
  {Keiser}},\ }\href@noop {} {\bibfield  {journal} {\bibinfo  {journal} {Wiley
  Encyclopedia of Telecommunications}\ } (\bibinfo {year} {2003})}\BibitemShut
  {NoStop}%
\bibitem [{\citenamefont {Mandel}\ and\ \citenamefont
  {Wolf}(1995)}]{mandel1995optical}%
  \BibitemOpen
  \bibfield  {author} {\bibinfo {author} {\bibfnamefont {L.}~\bibnamefont
  {Mandel}}\ and\ \bibinfo {author} {\bibfnamefont {E.}~\bibnamefont {Wolf}},\
  }\href@noop {} {\emph {\bibinfo {title} {Optical coherence and quantum
  optics}}}\ (\bibinfo  {publisher} {Cambridge university press},\ \bibinfo
  {year} {1995})\BibitemShut {NoStop}%
\bibitem [{\citenamefont {Zhao}\ and\ \citenamefont
  {Al{\`u}}(2011)}]{zhao2011manipulating}%
  \BibitemOpen
  \bibfield  {author} {\bibinfo {author} {\bibfnamefont {Y.}~\bibnamefont
  {Zhao}}\ and\ \bibinfo {author} {\bibfnamefont {A.}~\bibnamefont {Al{\`u}}},\
  }\href@noop {} {\bibfield  {journal} {\bibinfo  {journal} {Physical Review
  B}\ }\textbf {\bibinfo {volume} {84}},\ \bibinfo {pages} {205428} (\bibinfo
  {year} {2011})}\BibitemShut {NoStop}%
\bibitem [{\citenamefont {Glybovski}\ \emph {et~al.}(2016)\citenamefont
  {Glybovski}, \citenamefont {Tretyakov}, \citenamefont {Belov}, \citenamefont
  {Kivshar},\ and\ \citenamefont {Simovski}}]{glybovski2016metasurfaces}%
  \BibitemOpen
  \bibfield  {author} {\bibinfo {author} {\bibfnamefont {S.~B.}\ \bibnamefont
  {Glybovski}}, \bibinfo {author} {\bibfnamefont {S.~A.}\ \bibnamefont
  {Tretyakov}}, \bibinfo {author} {\bibfnamefont {P.~A.}\ \bibnamefont
  {Belov}}, \bibinfo {author} {\bibfnamefont {Y.~S.}\ \bibnamefont {Kivshar}},
  \ and\ \bibinfo {author} {\bibfnamefont {C.~R.}\ \bibnamefont {Simovski}},\
  }\href@noop {} {\bibfield  {journal} {\bibinfo  {journal} {Physics reports}\
  }\textbf {\bibinfo {volume} {634}},\ \bibinfo {pages} {1} (\bibinfo {year}
  {2016})}\BibitemShut {NoStop}%
\bibitem [{\citenamefont {Kruk}\ \emph {et~al.}(2016)\citenamefont {Kruk},
  \citenamefont {Hopkins}, \citenamefont {Kravchenko}, \citenamefont
  {Miroshnichenko}, \citenamefont {Neshev},\ and\ \citenamefont
  {Kivshar}}]{kruk2016invited}%
  \BibitemOpen
  \bibfield  {author} {\bibinfo {author} {\bibfnamefont {S.}~\bibnamefont
  {Kruk}}, \bibinfo {author} {\bibfnamefont {B.}~\bibnamefont {Hopkins}},
  \bibinfo {author} {\bibfnamefont {I.~I.}\ \bibnamefont {Kravchenko}},
  \bibinfo {author} {\bibfnamefont {A.}~\bibnamefont {Miroshnichenko}},
  \bibinfo {author} {\bibfnamefont {D.~N.}\ \bibnamefont {Neshev}}, \ and\
  \bibinfo {author} {\bibfnamefont {Y.~S.}\ \bibnamefont {Kivshar}},\
  }\href@noop {} {\bibfield  {journal} {\bibinfo  {journal} {Apl Photonics}\
  }\textbf {\bibinfo {volume} {1}},\ \bibinfo {pages} {030801} (\bibinfo {year}
  {2016})}\BibitemShut {NoStop}%
\bibitem [{\citenamefont {Kruk}\ and\ \citenamefont
  {Kivshar}(2017)}]{kruk2017functional}%
  \BibitemOpen
  \bibfield  {author} {\bibinfo {author} {\bibfnamefont {S.}~\bibnamefont
  {Kruk}}\ and\ \bibinfo {author} {\bibfnamefont {Y.}~\bibnamefont {Kivshar}},\
  }\href@noop {} {\bibfield  {journal} {\bibinfo  {journal} {ACS Photonics}\
  }\textbf {\bibinfo {volume} {4}},\ \bibinfo {pages} {2638} (\bibinfo {year}
  {2017})}\BibitemShut {NoStop}%
\bibitem [{\citenamefont {Lobanov}\ \emph {et~al.}(2015)\citenamefont
  {Lobanov}, \citenamefont {Weiss}, \citenamefont {Gippius}, \citenamefont
  {Tikhodeev}, \citenamefont {Kulakovskii}, \citenamefont {Konishi},\ and\
  \citenamefont {Kuwata-Gonokami}}]{lobanov2015polarization}%
  \BibitemOpen
  \bibfield  {author} {\bibinfo {author} {\bibfnamefont {S.~V.}\ \bibnamefont
  {Lobanov}}, \bibinfo {author} {\bibfnamefont {T.}~\bibnamefont {Weiss}},
  \bibinfo {author} {\bibfnamefont {N.~A.}\ \bibnamefont {Gippius}}, \bibinfo
  {author} {\bibfnamefont {S.~G.}\ \bibnamefont {Tikhodeev}}, \bibinfo {author}
  {\bibfnamefont {V.~D.}\ \bibnamefont {Kulakovskii}}, \bibinfo {author}
  {\bibfnamefont {K.}~\bibnamefont {Konishi}}, \ and\ \bibinfo {author}
  {\bibfnamefont {M.}~\bibnamefont {Kuwata-Gonokami}},\ }\href@noop {}
  {\bibfield  {journal} {\bibinfo  {journal} {Optics letters}\ }\textbf
  {\bibinfo {volume} {40}},\ \bibinfo {pages} {1528} (\bibinfo {year}
  {2015})}\BibitemShut {NoStop}%
\bibitem [{\citenamefont {Hsu}\ \emph {et~al.}(2017)\citenamefont {Hsu},
  \citenamefont {Zhen}, \citenamefont {Solja{\v{c}}i{\'c}},\ and\ \citenamefont
  {Stone}}]{hsu2017polarization}%
  \BibitemOpen
  \bibfield  {author} {\bibinfo {author} {\bibfnamefont {C.~W.}\ \bibnamefont
  {Hsu}}, \bibinfo {author} {\bibfnamefont {B.}~\bibnamefont {Zhen}}, \bibinfo
  {author} {\bibfnamefont {M.}~\bibnamefont {Solja{\v{c}}i{\'c}}}, \ and\
  \bibinfo {author} {\bibfnamefont {A.~D.}\ \bibnamefont {Stone}},\ }\href@noop
  {} {\bibfield  {journal} {\bibinfo  {journal} {arXiv preprint
  arXiv:1708.02197}\ } (\bibinfo {year} {2017})}\BibitemShut {NoStop}%
\bibitem [{\citenamefont {Guo}\ \emph {et~al.}(2017)\citenamefont {Guo},
  \citenamefont {Xiao},\ and\ \citenamefont {Fan}}]{guo2017topologically}%
  \BibitemOpen
  \bibfield  {author} {\bibinfo {author} {\bibfnamefont {Y.}~\bibnamefont
  {Guo}}, \bibinfo {author} {\bibfnamefont {M.}~\bibnamefont {Xiao}}, \ and\
  \bibinfo {author} {\bibfnamefont {S.}~\bibnamefont {Fan}},\ }\href@noop {}
  {\bibfield  {journal} {\bibinfo  {journal} {Physical review letters}\
  }\textbf {\bibinfo {volume} {119}},\ \bibinfo {pages} {167401} (\bibinfo
  {year} {2017})}\BibitemShut {NoStop}%
\bibitem [{\citenamefont {Guo}\ \emph {et~al.}(2019)\citenamefont {Guo},
  \citenamefont {Xiao}, \citenamefont {Zhou},\ and\ \citenamefont
  {Fan}}]{guo2019arbitrary}%
  \BibitemOpen
  \bibfield  {author} {\bibinfo {author} {\bibfnamefont {Y.}~\bibnamefont
  {Guo}}, \bibinfo {author} {\bibfnamefont {M.}~\bibnamefont {Xiao}}, \bibinfo
  {author} {\bibfnamefont {Y.}~\bibnamefont {Zhou}}, \ and\ \bibinfo {author}
  {\bibfnamefont {S.}~\bibnamefont {Fan}},\ }\href@noop {} {\bibfield
  {journal} {\bibinfo  {journal} {Advanced Optical Materials}\ ,\ \bibinfo
  {pages} {1801453}} (\bibinfo {year} {2019})}\BibitemShut {NoStop}%
\bibitem [{\citenamefont {Hsu}\ \emph {et~al.}(2013)\citenamefont {Hsu},
  \citenamefont {Zhen}, \citenamefont {Lee}, \citenamefont {Chua},
  \citenamefont {Johnson}, \citenamefont {Joannopoulos},\ and\ \citenamefont
  {Solja{\v{c}}i{\'c}}}]{hsu2013observation}%
  \BibitemOpen
  \bibfield  {author} {\bibinfo {author} {\bibfnamefont {C.~W.}\ \bibnamefont
  {Hsu}}, \bibinfo {author} {\bibfnamefont {B.}~\bibnamefont {Zhen}}, \bibinfo
  {author} {\bibfnamefont {J.}~\bibnamefont {Lee}}, \bibinfo {author}
  {\bibfnamefont {S.-L.}\ \bibnamefont {Chua}}, \bibinfo {author}
  {\bibfnamefont {S.~G.}\ \bibnamefont {Johnson}}, \bibinfo {author}
  {\bibfnamefont {J.~D.}\ \bibnamefont {Joannopoulos}}, \ and\ \bibinfo
  {author} {\bibfnamefont {M.}~\bibnamefont {Solja{\v{c}}i{\'c}}},\ }\href@noop
  {} {\bibfield  {journal} {\bibinfo  {journal} {Nature}\ }\textbf {\bibinfo
  {volume} {499}},\ \bibinfo {pages} {188} (\bibinfo {year}
  {2013})}\BibitemShut {NoStop}%
\bibitem [{\citenamefont {Zhen}\ \emph {et~al.}(2014)\citenamefont {Zhen},
  \citenamefont {Hsu}, \citenamefont {Lu}, \citenamefont {Stone},\ and\
  \citenamefont {Solja{\v{c}}i{\'c}}}]{zhen2014topological}%
  \BibitemOpen
  \bibfield  {author} {\bibinfo {author} {\bibfnamefont {B.}~\bibnamefont
  {Zhen}}, \bibinfo {author} {\bibfnamefont {C.~W.}\ \bibnamefont {Hsu}},
  \bibinfo {author} {\bibfnamefont {L.}~\bibnamefont {Lu}}, \bibinfo {author}
  {\bibfnamefont {A.~D.}\ \bibnamefont {Stone}}, \ and\ \bibinfo {author}
  {\bibfnamefont {M.}~\bibnamefont {Solja{\v{c}}i{\'c}}},\ }\href@noop {}
  {\bibfield  {journal} {\bibinfo  {journal} {Physical review letters}\
  }\textbf {\bibinfo {volume} {113}},\ \bibinfo {pages} {257401} (\bibinfo
  {year} {2014})}\BibitemShut {NoStop}%
\bibitem [{\citenamefont {Yang}\ \emph {et~al.}(2014)\citenamefont {Yang},
  \citenamefont {Peng}, \citenamefont {Liang}, \citenamefont {Li},\ and\
  \citenamefont {Noda}}]{yang2014analytical}%
  \BibitemOpen
  \bibfield  {author} {\bibinfo {author} {\bibfnamefont {Y.}~\bibnamefont
  {Yang}}, \bibinfo {author} {\bibfnamefont {C.}~\bibnamefont {Peng}}, \bibinfo
  {author} {\bibfnamefont {Y.}~\bibnamefont {Liang}}, \bibinfo {author}
  {\bibfnamefont {Z.}~\bibnamefont {Li}}, \ and\ \bibinfo {author}
  {\bibfnamefont {S.}~\bibnamefont {Noda}},\ }\href@noop {} {\bibfield
  {journal} {\bibinfo  {journal} {Physical review letters}\ }\textbf {\bibinfo
  {volume} {113}},\ \bibinfo {pages} {037401} (\bibinfo {year}
  {2014})}\BibitemShut {NoStop}%
\bibitem [{\citenamefont {Bulgakov}\ and\ \citenamefont
  {Maksimov}(2017)}]{bulgakov2017bound}%
  \BibitemOpen
  \bibfield  {author} {\bibinfo {author} {\bibfnamefont {E.~N.}\ \bibnamefont
  {Bulgakov}}\ and\ \bibinfo {author} {\bibfnamefont {D.~N.}\ \bibnamefont
  {Maksimov}},\ }\href@noop {} {\bibfield  {journal} {\bibinfo  {journal}
  {Physical Review A}\ }\textbf {\bibinfo {volume} {96}},\ \bibinfo {pages}
  {063833} (\bibinfo {year} {2017})}\BibitemShut {NoStop}%
\bibitem [{\citenamefont {Zhang}\ \emph {et~al.}(2018)\citenamefont {Zhang},
  \citenamefont {Chen}, \citenamefont {Liu}, \citenamefont {Hsu}, \citenamefont
  {Wang}, \citenamefont {Guan}, \citenamefont {Liu}, \citenamefont {Shi},
  \citenamefont {Lu},\ and\ \citenamefont {Zi}}]{zhang2018observation}%
  \BibitemOpen
  \bibfield  {author} {\bibinfo {author} {\bibfnamefont {Y.}~\bibnamefont
  {Zhang}}, \bibinfo {author} {\bibfnamefont {A.}~\bibnamefont {Chen}},
  \bibinfo {author} {\bibfnamefont {W.}~\bibnamefont {Liu}}, \bibinfo {author}
  {\bibfnamefont {C.~W.}\ \bibnamefont {Hsu}}, \bibinfo {author} {\bibfnamefont
  {B.}~\bibnamefont {Wang}}, \bibinfo {author} {\bibfnamefont {F.}~\bibnamefont
  {Guan}}, \bibinfo {author} {\bibfnamefont {X.}~\bibnamefont {Liu}}, \bibinfo
  {author} {\bibfnamefont {L.}~\bibnamefont {Shi}}, \bibinfo {author}
  {\bibfnamefont {L.}~\bibnamefont {Lu}}, \ and\ \bibinfo {author}
  {\bibfnamefont {J.}~\bibnamefont {Zi}},\ }\href@noop {} {\bibfield  {journal}
  {\bibinfo  {journal} {Physical review letters}\ }\textbf {\bibinfo {volume}
  {120}},\ \bibinfo {pages} {186103} (\bibinfo {year} {2018})}\BibitemShut
  {NoStop}%
\bibitem [{\citenamefont {Doeleman}\ \emph {et~al.}(2018)\citenamefont
  {Doeleman}, \citenamefont {Monticone}, \citenamefont {den Hollander},
  \citenamefont {Al{\`u}},\ and\ \citenamefont
  {Koenderink}}]{doeleman2018experimental}%
  \BibitemOpen
  \bibfield  {author} {\bibinfo {author} {\bibfnamefont {H.~M.}\ \bibnamefont
  {Doeleman}}, \bibinfo {author} {\bibfnamefont {F.}~\bibnamefont {Monticone}},
  \bibinfo {author} {\bibfnamefont {W.}~\bibnamefont {den Hollander}}, \bibinfo
  {author} {\bibfnamefont {A.}~\bibnamefont {Al{\`u}}}, \ and\ \bibinfo
  {author} {\bibfnamefont {A.~F.}\ \bibnamefont {Koenderink}},\ }\href@noop {}
  {\bibfield  {journal} {\bibinfo  {journal} {Nature Photonics}\ }\textbf
  {\bibinfo {volume} {12}},\ \bibinfo {pages} {397} (\bibinfo {year}
  {2018})}\BibitemShut {NoStop}%
\bibitem [{\citenamefont {Song}\ \emph
  {et~al.}(2018{\natexlab{a}})\citenamefont {Song}, \citenamefont {Jiang},
  \citenamefont {Liu}, \citenamefont {Hu},\ and\ \citenamefont
  {Zi}}]{song2018cherenkov}%
  \BibitemOpen
  \bibfield  {author} {\bibinfo {author} {\bibfnamefont {Y.}~\bibnamefont
  {Song}}, \bibinfo {author} {\bibfnamefont {N.}~\bibnamefont {Jiang}},
  \bibinfo {author} {\bibfnamefont {L.}~\bibnamefont {Liu}}, \bibinfo {author}
  {\bibfnamefont {X.}~\bibnamefont {Hu}}, \ and\ \bibinfo {author}
  {\bibfnamefont {J.}~\bibnamefont {Zi}},\ }\href@noop {} {\bibfield  {journal}
  {\bibinfo  {journal} {Physical Review Applied}\ }\textbf {\bibinfo {volume}
  {10}},\ \bibinfo {pages} {064026} (\bibinfo {year}
  {2018}{\natexlab{a}})}\BibitemShut {NoStop}%
\bibitem [{\citenamefont {Dai}\ \emph {et~al.}(2018)\citenamefont {Dai},
  \citenamefont {Liu}, \citenamefont {Han},\ and\ \citenamefont
  {Zi}}]{dai2018topologically}%
  \BibitemOpen
  \bibfield  {author} {\bibinfo {author} {\bibfnamefont {S.}~\bibnamefont
  {Dai}}, \bibinfo {author} {\bibfnamefont {L.}~\bibnamefont {Liu}}, \bibinfo
  {author} {\bibfnamefont {D.}~\bibnamefont {Han}}, \ and\ \bibinfo {author}
  {\bibfnamefont {J.}~\bibnamefont {Zi}},\ }\href@noop {} {\bibfield  {journal}
  {\bibinfo  {journal} {Physical Review B}\ }\textbf {\bibinfo {volume} {98}},\
  \bibinfo {pages} {081405} (\bibinfo {year} {2018})}\BibitemShut {NoStop}%
\bibitem [{\citenamefont {He}\ \emph {et~al.}(2018)\citenamefont {He},
  \citenamefont {Guo}, \citenamefont {Feng}, \citenamefont {Xu},\ and\
  \citenamefont {Miroshnichenko}}]{he2018toroidal}%
  \BibitemOpen
  \bibfield  {author} {\bibinfo {author} {\bibfnamefont {Y.}~\bibnamefont
  {He}}, \bibinfo {author} {\bibfnamefont {G.}~\bibnamefont {Guo}}, \bibinfo
  {author} {\bibfnamefont {T.}~\bibnamefont {Feng}}, \bibinfo {author}
  {\bibfnamefont {Y.}~\bibnamefont {Xu}}, \ and\ \bibinfo {author}
  {\bibfnamefont {A.~E.}\ \bibnamefont {Miroshnichenko}},\ }\href@noop {}
  {\bibfield  {journal} {\bibinfo  {journal} {Physical Review B}\ }\textbf
  {\bibinfo {volume} {98}},\ \bibinfo {pages} {161112} (\bibinfo {year}
  {2018})}\BibitemShut {NoStop}%
\bibitem [{\citenamefont {Jin}\ \emph {et~al.}(2018)\citenamefont {Jin},
  \citenamefont {Yin}, \citenamefont {Ni}, \citenamefont {Solja{\v{c}}i{\'c}},
  \citenamefont {Zhen},\ and\ \citenamefont {Peng}}]{jin2018topologically}%
  \BibitemOpen
  \bibfield  {author} {\bibinfo {author} {\bibfnamefont {J.}~\bibnamefont
  {Jin}}, \bibinfo {author} {\bibfnamefont {X.}~\bibnamefont {Yin}}, \bibinfo
  {author} {\bibfnamefont {L.}~\bibnamefont {Ni}}, \bibinfo {author}
  {\bibfnamefont {M.}~\bibnamefont {Solja{\v{c}}i{\'c}}}, \bibinfo {author}
  {\bibfnamefont {B.}~\bibnamefont {Zhen}}, \ and\ \bibinfo {author}
  {\bibfnamefont {C.}~\bibnamefont {Peng}},\ }\href@noop {} {\bibfield
  {journal} {\bibinfo  {journal} {arXiv preprint arXiv:1812.00892}\ } (\bibinfo
  {year} {2018})}\BibitemShut {NoStop}%
\bibitem [{\citenamefont {Koshelev}\ \emph {et~al.}(2018)\citenamefont
  {Koshelev}, \citenamefont {Lepeshov}, \citenamefont {Liu}, \citenamefont
  {Bogdanov},\ and\ \citenamefont {Kivshar}}]{koshelev2018asymmetric}%
  \BibitemOpen
  \bibfield  {author} {\bibinfo {author} {\bibfnamefont {K.}~\bibnamefont
  {Koshelev}}, \bibinfo {author} {\bibfnamefont {S.}~\bibnamefont {Lepeshov}},
  \bibinfo {author} {\bibfnamefont {M.}~\bibnamefont {Liu}}, \bibinfo {author}
  {\bibfnamefont {A.}~\bibnamefont {Bogdanov}}, \ and\ \bibinfo {author}
  {\bibfnamefont {Y.}~\bibnamefont {Kivshar}},\ }\href@noop {} {\bibfield
  {journal} {\bibinfo  {journal} {Physical review letters}\ }\textbf {\bibinfo
  {volume} {121}},\ \bibinfo {pages} {193903} (\bibinfo {year}
  {2018})}\BibitemShut {NoStop}%
\bibitem [{\citenamefont {Chen}\ \emph {et~al.}(2019)\citenamefont {Chen},
  \citenamefont {Chen},\ and\ \citenamefont {Liu}}]{chen2019singularities}%
  \BibitemOpen
  \bibfield  {author} {\bibinfo {author} {\bibfnamefont {W.}~\bibnamefont
  {Chen}}, \bibinfo {author} {\bibfnamefont {Y.}~\bibnamefont {Chen}}, \ and\
  \bibinfo {author} {\bibfnamefont {W.}~\bibnamefont {Liu}},\ }\href@noop {}
  {\bibfield  {journal} {\bibinfo  {journal} {arXiv preprint arXiv:1901.04159}\
  } (\bibinfo {year} {2019})}\BibitemShut {NoStop}%
\bibitem [{\citenamefont {Cerjan}\ \emph {et~al.}(2019)\citenamefont {Cerjan},
  \citenamefont {Hsu},\ and\ \citenamefont {Rechtsman}}]{cerjan2019bound}%
  \BibitemOpen
  \bibfield  {author} {\bibinfo {author} {\bibfnamefont {A.}~\bibnamefont
  {Cerjan}}, \bibinfo {author} {\bibfnamefont {C.~W.}\ \bibnamefont {Hsu}}, \
  and\ \bibinfo {author} {\bibfnamefont {M.~C.}\ \bibnamefont {Rechtsman}},\
  }\href@noop {} {\bibfield  {journal} {\bibinfo  {journal} {arXiv preprint
  arXiv:1901.07126}\ } (\bibinfo {year} {2019})}\BibitemShut {NoStop}%
\bibitem [{\citenamefont {Koshelev}\ \emph {et~al.}(2019)\citenamefont
  {Koshelev}, \citenamefont {Favraud}, \citenamefont {Bogdanov}, \citenamefont
  {Kivshar},\ and\ \citenamefont {Fratalocchi}}]{koshelev2019nonradiating}%
  \BibitemOpen
  \bibfield  {author} {\bibinfo {author} {\bibfnamefont {K.}~\bibnamefont
  {Koshelev}}, \bibinfo {author} {\bibfnamefont {G.}~\bibnamefont {Favraud}},
  \bibinfo {author} {\bibfnamefont {A.}~\bibnamefont {Bogdanov}}, \bibinfo
  {author} {\bibfnamefont {Y.}~\bibnamefont {Kivshar}}, \ and\ \bibinfo
  {author} {\bibfnamefont {A.}~\bibnamefont {Fratalocchi}},\ }\href@noop {}
  {\bibfield  {journal} {\bibinfo  {journal} {arXiv preprint arXiv:1903.04756}\
  } (\bibinfo {year} {2019})}\BibitemShut {NoStop}%
\bibitem [{\citenamefont {Sadrieva}\ \emph {et~al.}(2019)\citenamefont
  {Sadrieva}, \citenamefont {Frizyuk}, \citenamefont {Petrov}, \citenamefont
  {Kivshar},\ and\ \citenamefont {Bogdanov}}]{sadrieva2019multipolar}%
  \BibitemOpen
  \bibfield  {author} {\bibinfo {author} {\bibfnamefont {Z.}~\bibnamefont
  {Sadrieva}}, \bibinfo {author} {\bibfnamefont {K.}~\bibnamefont {Frizyuk}},
  \bibinfo {author} {\bibfnamefont {M.}~\bibnamefont {Petrov}}, \bibinfo
  {author} {\bibfnamefont {Y.}~\bibnamefont {Kivshar}}, \ and\ \bibinfo
  {author} {\bibfnamefont {A.}~\bibnamefont {Bogdanov}},\ }\href@noop {}
  {\bibfield  {journal} {\bibinfo  {journal} {arXiv preprint arXiv:1903.00309}\
  } (\bibinfo {year} {2019})}\BibitemShut {NoStop}%
\bibitem [{\citenamefont {Zhou}\ \emph {et~al.}(2018)\citenamefont {Zhou},
  \citenamefont {Peng}, \citenamefont {Yoon}, \citenamefont {Hsu},
  \citenamefont {Nelson}, \citenamefont {Fu}, \citenamefont {Joannopoulos},
  \citenamefont {Solja{\v{c}}i{\'c}},\ and\ \citenamefont
  {Zhen}}]{zhou2018observation}%
  \BibitemOpen
  \bibfield  {author} {\bibinfo {author} {\bibfnamefont {H.}~\bibnamefont
  {Zhou}}, \bibinfo {author} {\bibfnamefont {C.}~\bibnamefont {Peng}}, \bibinfo
  {author} {\bibfnamefont {Y.}~\bibnamefont {Yoon}}, \bibinfo {author}
  {\bibfnamefont {C.~W.}\ \bibnamefont {Hsu}}, \bibinfo {author} {\bibfnamefont
  {K.~A.}\ \bibnamefont {Nelson}}, \bibinfo {author} {\bibfnamefont
  {L.}~\bibnamefont {Fu}}, \bibinfo {author} {\bibfnamefont {J.~D.}\
  \bibnamefont {Joannopoulos}}, \bibinfo {author} {\bibfnamefont
  {M.}~\bibnamefont {Solja{\v{c}}i{\'c}}}, \ and\ \bibinfo {author}
  {\bibfnamefont {B.}~\bibnamefont {Zhen}},\ }\href@noop {} {\bibfield
  {journal} {\bibinfo  {journal} {Science}\ }\textbf {\bibinfo {volume}
  {359}},\ \bibinfo {pages} {1009} (\bibinfo {year} {2018})}\BibitemShut
  {NoStop}%
\bibitem [{\citenamefont {Nye}(1999)}]{nye1999natural}%
  \BibitemOpen
  \bibfield  {author} {\bibinfo {author} {\bibfnamefont {J.~F.}\ \bibnamefont
  {Nye}},\ }\href@noop {} {\emph {\bibinfo {title} {Natural focusing and fine
  structure of light: caustics and wave dislocations}}}\ (\bibinfo  {publisher}
  {CRC Press},\ \bibinfo {year} {1999})\BibitemShut {NoStop}%
\bibitem [{\citenamefont {Berry}\ and\ \citenamefont
  {Dennis}(2001)}]{berry2001polarization}%
  \BibitemOpen
  \bibfield  {author} {\bibinfo {author} {\bibfnamefont {M.}~\bibnamefont
  {Berry}}\ and\ \bibinfo {author} {\bibfnamefont {M.}~\bibnamefont {Dennis}},\
  }\href@noop {} {\bibfield  {journal} {\bibinfo  {journal} {Proceedings:
  Mathematics, Physical and Engineering Sciences}\ ,\ \bibinfo {pages} {141}}
  (\bibinfo {year} {2001})}\BibitemShut {NoStop}%
\bibitem [{\citenamefont {Dennis}(2002)}]{dennis2002polarization}%
  \BibitemOpen
  \bibfield  {author} {\bibinfo {author} {\bibfnamefont {M.}~\bibnamefont
  {Dennis}},\ }\href@noop {} {\bibfield  {journal} {\bibinfo  {journal} {Optics
  Communications}\ }\textbf {\bibinfo {volume} {213}},\ \bibinfo {pages} {201}
  (\bibinfo {year} {2002})}\BibitemShut {NoStop}%
\bibitem [{\citenamefont {Freund}(2002)}]{freund2002polarization}%
  \BibitemOpen
  \bibfield  {author} {\bibinfo {author} {\bibfnamefont {I.}~\bibnamefont
  {Freund}},\ }\href@noop {} {\bibfield  {journal} {\bibinfo  {journal} {Optics
  communications}\ }\textbf {\bibinfo {volume} {201}},\ \bibinfo {pages} {251}
  (\bibinfo {year} {2002})}\BibitemShut {NoStop}%
\bibitem [{\citenamefont {Schoonover}\ and\ \citenamefont
  {Visser}(2006)}]{schoonover2006polarization}%
  \BibitemOpen
  \bibfield  {author} {\bibinfo {author} {\bibfnamefont {R.~W.}\ \bibnamefont
  {Schoonover}}\ and\ \bibinfo {author} {\bibfnamefont {T.~D.}\ \bibnamefont
  {Visser}},\ }\href@noop {} {\bibfield  {journal} {\bibinfo  {journal} {Optics
  Express}\ }\textbf {\bibinfo {volume} {14}},\ \bibinfo {pages} {5733}
  (\bibinfo {year} {2006})}\BibitemShut {NoStop}%
\bibitem [{\citenamefont {Burresi}\ \emph {et~al.}(2009)\citenamefont
  {Burresi}, \citenamefont {Engelen}, \citenamefont {Opheij}, \citenamefont
  {van Oosten}, \citenamefont {Mori}, \citenamefont {Baba},\ and\ \citenamefont
  {Kuipers}}]{burresi2009observation}%
  \BibitemOpen
  \bibfield  {author} {\bibinfo {author} {\bibfnamefont {M.}~\bibnamefont
  {Burresi}}, \bibinfo {author} {\bibfnamefont {R.}~\bibnamefont {Engelen}},
  \bibinfo {author} {\bibfnamefont {A.}~\bibnamefont {Opheij}}, \bibinfo
  {author} {\bibfnamefont {D.}~\bibnamefont {van Oosten}}, \bibinfo {author}
  {\bibfnamefont {D.}~\bibnamefont {Mori}}, \bibinfo {author} {\bibfnamefont
  {T.}~\bibnamefont {Baba}}, \ and\ \bibinfo {author} {\bibfnamefont
  {L.}~\bibnamefont {Kuipers}},\ }\href@noop {} {\bibfield  {journal} {\bibinfo
   {journal} {Physical review letters}\ }\textbf {\bibinfo {volume} {102}},\
  \bibinfo {pages} {033902} (\bibinfo {year} {2009})}\BibitemShut {NoStop}%
\bibitem [{\citenamefont {Song}\ \emph
  {et~al.}(2018{\natexlab{b}})\citenamefont {Song}, \citenamefont {Leykam},
  \citenamefont {Su}, \citenamefont {Liu}, \citenamefont {Tang}, \citenamefont
  {Liu}, \citenamefont {Zhao}, \citenamefont {Efremidis}, \citenamefont {Xu},\
  and\ \citenamefont {Chen}}]{song2018valley}%
  \BibitemOpen
  \bibfield  {author} {\bibinfo {author} {\bibfnamefont {D.}~\bibnamefont
  {Song}}, \bibinfo {author} {\bibfnamefont {D.}~\bibnamefont {Leykam}},
  \bibinfo {author} {\bibfnamefont {J.}~\bibnamefont {Su}}, \bibinfo {author}
  {\bibfnamefont {X.}~\bibnamefont {Liu}}, \bibinfo {author} {\bibfnamefont
  {L.}~\bibnamefont {Tang}}, \bibinfo {author} {\bibfnamefont {S.}~\bibnamefont
  {Liu}}, \bibinfo {author} {\bibfnamefont {J.}~\bibnamefont {Zhao}}, \bibinfo
  {author} {\bibfnamefont {N.~K.}\ \bibnamefont {Efremidis}}, \bibinfo {author}
  {\bibfnamefont {J.}~\bibnamefont {Xu}}, \ and\ \bibinfo {author}
  {\bibfnamefont {Z.}~\bibnamefont {Chen}},\ }\href@noop {} {\bibfield
  {journal} {\bibinfo  {journal} {arXiv preprint arXiv:1810.12736}\ } (\bibinfo
  {year} {2018}{\natexlab{b}})}\BibitemShut {NoStop}%
\bibitem [{\citenamefont {Bliokh}\ \emph {et~al.}(2019)\citenamefont {Bliokh},
  \citenamefont {Alonso},\ and\ \citenamefont {Dennis}}]{bliokh2019geometric}%
  \BibitemOpen
  \bibfield  {author} {\bibinfo {author} {\bibfnamefont {K.~Y.}\ \bibnamefont
  {Bliokh}}, \bibinfo {author} {\bibfnamefont {M.~A.}\ \bibnamefont {Alonso}},
  \ and\ \bibinfo {author} {\bibfnamefont {M.~R.}\ \bibnamefont {Dennis}},\
  }\href@noop {} {\bibfield  {journal} {\bibinfo  {journal} {arXiv preprint
  arXiv:1903.01304}\ } (\bibinfo {year} {2019})}\BibitemShut {NoStop}%
\bibitem [{\citenamefont {D’Errico}\ \emph {et~al.}(2017)\citenamefont
  {D’Errico}, \citenamefont {Maffei}, \citenamefont {Piccirillo},
  \citenamefont {De~Lisio}, \citenamefont {Cardano},\ and\ \citenamefont
  {Marrucci}}]{d2017topological}%
  \BibitemOpen
  \bibfield  {author} {\bibinfo {author} {\bibfnamefont {A.}~\bibnamefont
  {D’Errico}}, \bibinfo {author} {\bibfnamefont {M.}~\bibnamefont {Maffei}},
  \bibinfo {author} {\bibfnamefont {B.}~\bibnamefont {Piccirillo}}, \bibinfo
  {author} {\bibfnamefont {C.}~\bibnamefont {De~Lisio}}, \bibinfo {author}
  {\bibfnamefont {F.}~\bibnamefont {Cardano}}, \ and\ \bibinfo {author}
  {\bibfnamefont {L.}~\bibnamefont {Marrucci}},\ }\href@noop {} {\bibfield
  {journal} {\bibinfo  {journal} {Scientific reports}\ }\textbf {\bibinfo
  {volume} {7}},\ \bibinfo {pages} {40195} (\bibinfo {year}
  {2017})}\BibitemShut {NoStop}%
\bibitem [{\citenamefont {Otte}\ \emph {et~al.}(2018)\citenamefont {Otte},
  \citenamefont {Alpmann},\ and\ \citenamefont {Denz}}]{otte2018polarization}%
  \BibitemOpen
  \bibfield  {author} {\bibinfo {author} {\bibfnamefont {E.}~\bibnamefont
  {Otte}}, \bibinfo {author} {\bibfnamefont {C.}~\bibnamefont {Alpmann}}, \
  and\ \bibinfo {author} {\bibfnamefont {C.}~\bibnamefont {Denz}},\ }\href@noop
  {} {\bibfield  {journal} {\bibinfo  {journal} {Laser \& Photonics Reviews}\
  }\textbf {\bibinfo {volume} {12}},\ \bibinfo {pages} {1700200} (\bibinfo
  {year} {2018})}\BibitemShut {NoStop}%
\bibitem [{sm()}]{sm}%
  \BibitemOpen
  \href@noop {} {}\bibinfo {note} {See Supplementary Material.}\BibitemShut
  {Stop}%
\bibitem [{\citenamefont {Yin}\ \emph {et~al.}(2013)\citenamefont {Yin},
  \citenamefont {Ye}, \citenamefont {Rho}, \citenamefont {Wang},\ and\
  \citenamefont {Zhang}}]{yin2013photonic}%
  \BibitemOpen
  \bibfield  {author} {\bibinfo {author} {\bibfnamefont {X.}~\bibnamefont
  {Yin}}, \bibinfo {author} {\bibfnamefont {Z.}~\bibnamefont {Ye}}, \bibinfo
  {author} {\bibfnamefont {J.}~\bibnamefont {Rho}}, \bibinfo {author}
  {\bibfnamefont {Y.}~\bibnamefont {Wang}}, \ and\ \bibinfo {author}
  {\bibfnamefont {X.}~\bibnamefont {Zhang}},\ }\href@noop {} {\bibfield
  {journal} {\bibinfo  {journal} {Science}\ }\textbf {\bibinfo {volume}
  {339}},\ \bibinfo {pages} {1405} (\bibinfo {year} {2013})}\BibitemShut
  {NoStop}%
\bibitem [{\citenamefont {Bliokh}\ \emph {et~al.}(2015)\citenamefont {Bliokh},
  \citenamefont {Rodr{\'\i}guez-Fortu{\~n}o}, \citenamefont {Nori},\ and\
  \citenamefont {Zayats}}]{bliokh2015spin}%
  \BibitemOpen
  \bibfield  {author} {\bibinfo {author} {\bibfnamefont {K.~Y.}\ \bibnamefont
  {Bliokh}}, \bibinfo {author} {\bibfnamefont {F.~J.}\ \bibnamefont
  {Rodr{\'\i}guez-Fortu{\~n}o}}, \bibinfo {author} {\bibfnamefont
  {F.}~\bibnamefont {Nori}}, \ and\ \bibinfo {author} {\bibfnamefont {A.~V.}\
  \bibnamefont {Zayats}},\ }\href@noop {} {\bibfield  {journal} {\bibinfo
  {journal} {Nature Photonics}\ }\textbf {\bibinfo {volume} {9}},\ \bibinfo
  {pages} {796} (\bibinfo {year} {2015})}\BibitemShut {NoStop}%
\end{thebibliography}

%

\end{document}